\documentclass[%
 reprint,
superscriptaddress,
 amsmath,amssymb,
 aps,
]{revtex4-2}

\usepackage{hyperref}
\usepackage{graphicx}
\usepackage{dcolumn}
\usepackage{bm}

\usepackage{setspace}
\usepackage[capitalise]{cleveref}
\usepackage{physics}
\usepackage{xspace}
\newcommand{\mv}[1]{\mathbf{#1}}
\newcommand{\lmax}{\ensuremath{l_{\text{max}}}}

\newcommand{\supprefc}{Supp.~Fig.~3\xspace}
\newcommand{\supprefd}{Supp.~Fig.~4\xspace}

\newcommand{\cs}{~}

\newcounter{myfigpanel}[figure]

\newcommand{\panelletter}[1]{\refstepcounter{myfigpanel}\label{#1}\Alph{myfigpanel}}
\newcommand{\panel}[1]{(\protect\panelletter{#1})}
\newcommand{\silentpanel}[1]{{\protect\refstepcounter{myfigpanel}\label{#1}}}
\makeatletter

\newcommand{\@fragilepanels}[3]{
    (\panelletter{#1}--
    \@for\@pref:={#2}\do{\silentpanel{\@pref}}%
    \panelletter{#3})
}%
\newcommand{\panels}{\protect\@fragilepanels}

\newcommand{\mycaption}[1]{\caption{\setcounter{myfigpanel}{0}#1}}

\makeatother
\usepackage{contour, upgreek} 
\contourlength{0.01em}
\newcommand{\bsigma}{\contour[auto]{black}{\ensuremath{\upsigma}}}
\newcommand{\bmu}{\contour[auto]{black}{\ensuremath{\upmu}}}

\newcommand{\MPIDS}{\affiliation{Max Planck Institute for Dynamics and Self-Organization, Göttingen, Germany}}
\newcommand{\IDCS}{\affiliation{Institute for the Dynamics of Complex Systems, Göttingen University, Göttingen, Germany}}
\newcommand{\RPCTP}{\affiliation{Rudolf Peierls Centre for Theoretical Physics, University of Oxford, Oxford OX1 3PU, United Kingdom}}

\begin{document}


\title{Stress anisotropy in
confined populations of growing rods}

\author{Jonas Isensee}
\MPIDS
\IDCS
\author{Lukas Hupe}%
\MPIDS
\IDCS
\author{Ramin Golestanian}
\MPIDS
\IDCS
\RPCTP
\author{Philip Bittihn}
\email{philip.bittihn@ds.mpg.de}
\MPIDS
\IDCS


\begin{abstract}
Order and alignment are ubiquitous in growing colonies of rod-shaped bacteria due to the nematic properties of the constituent particles.
These effects are the result of the active stresses generated by growth, passive mechanical interactions between cells, and flow-induced effects due to the shape of the confining container.
However, how these contributing factors interact to give rise to the observed global alignment patterns remains elusive.
Here, we study, \emph{in-silico}, colonies of growing rod-shaped particles of different aspect ratios confined in channel-like geometries.
A spatially resolved analysis of the stress tensor reveals a strong relationship between near-perfect alignment and an inversion of stress anisotropy for particles with large length-to-width ratios.
We show that, in quantitative agreement with an asymptotic theory, strong alignment can lead to a decoupling of active and passive stresses parallel and perpendicular to the direction of growth, respectively.
We demonstrate the robustness of these effects in a geometry that provides less restrictive confinement and introduces natural perturbations in alignment. 
Our results illustrate the complexity arising from the inherent coupling between nematic order and active stresses in growing active matter which is modulated by geometric and configurational constraints due to confinement.
\end{abstract}

\maketitle


Self-organization in multicellular biological systems is driven by inherent cellular activity.
This non-equilibrium activity can take a variety of forms, including self-propulsion from cell motility\cs\cite{HenkesMonolayers2020,GompperRoadmap2020}, active adhesion\cs\cite{kuanCellularAggregates2021} or chemical activity from metabolism or signaling\cs\cite{CarmonaTumor2017,TodaSignaling2018}, which also occur in non-biological active matter\cs\cite{golestanian2019_phoretic,LiebchenSynthetic2018}.
Growth is one of the hallmarks of life. It constitutes another process by which energy can be injected into the system at the microscopic scale, and has been shown to able to balance out chemical interactions at the level of large-scale behaviour \cite{gelimson2015_collective}.
The mechanisms by which \emph{growing active matter} self-organizes to form multicellular communities such as biofilms\cs\cite{FlemmingBiofilmReview2016, WongReviewBiofilms2021} and functioning tissues\cs\cite{BryantGrowthCoupling2016,MORITA2017354, ranftFluidization2010} are complex and often involve other forms of activity or internal regulation\cs\cite{Pollack2022deadmatter, CaoSynthetic2016, bittihn_genetically_2020,gelimson2016_multicellular}.
Here, we focus on the mechanical aspects of growth, mediated by steric interactions between individual rod-shaped cells and confinement. They are sufficient to reproduce alignment and large-scale flow patterns observed in the initial stages of bacterial colony formation\cs\cite{DoostmohammadiDefectGrowing2016,you_geometry_2018,Hartmann3DBiofilm2019,basaranOrientationalOrderInward2022} and represent a prime example for the class of systems known as active nematics\cs\cite{BalasubramaniamActiveNematics2022}.

The consequences of these ingredients depend critically on the mechanical environment:
For freely expanding colonies in two dimensions, the overall colony shows no preferred direction but generates locally ordered microdomains arising from a competition between passive elastic properties and active extensile stresses generated by growth, which are themselves functions of cell properties such as their length-to-width aspect ratio and growth rate\cs\cite{you_geometry_2018,dellarcipreteGrowingBacterialColony2018}. Topological defects typical for active nematics, which arise naturally at the domain boundaries\cs\cite{dellarcipreteGrowingBacterialColony2018,van_holthe_tot_echten_defect_2020}, have also been shown to be important for the transition into the third dimension\cs\cite{grantRoleMechanicalForces2014, copenhagenTopologicalDefectsPromote2021}.

In contrast, both in nature and lab experiments, colonies often grow in confined environments. A prominent example are rectangular channels, where populations of growing rod-shaped cells develop global nematic order, with their elongation direction oriented towards the exits (see Fig.~\ref{pan:experiment}). Despite the simple setup, these systems show a plethora of phenomena which have served to elucidate fundamental processes governing growing active nematics in a number of studies. In the course of these investigations, the emergence and maintenance of global orientational order were attributed to the response of the elongated particles to growth-induced expansion flow\cs\cite{volfson_biomechanical_2008} as well as to globally anisotropic stresses which build up due the distinct boundary conditions in different directions\cs\cite{youConfinementinducedSelforganizationGrowing2021}. Similar to the \emph{local} order characterized by the microdomain size in freely expanding colonies, the orientational order parameter for this \emph{global} alignment
 was found to depend on the mechanical parameters of the system, such as the length-to-width ratios of the growing rods and the growth rate, the latter also confirmed in experiments\cs\cite{sheatsRoleGrowthRate}. For near-perfect alignment, the system also becomes inhomogeneous in time and is characterized by intermittent breakdowns of order due to a buckling instability\cs\cite{boyerBucklingInstabilityOrdered2011,orozco-fuentesOrderIntermittencyPressure2013}.

The picture of anisotropic stress maintaining global order is appealing, as it can be directly related to the geometry of the system: In the direction of the openings, stress can be dissipated more easily than in the confined direction perpendicular to it and this perpendicular stress can help stabilize the emergent order, leading to a self-consistent, dynamic steady state. While this phenomenology was reproduced by a visco-elastic continuum theory and was also observed without the perpendicular confinement\cs\cite{youConfinementinducedSelforganizationGrowing2021}, the buckling analogy\cs\cite{boyerBucklingInstabilityOrdered2011} (which is also related to anisotropic stress) and resulting dynamical complexity\cs\cite{orozco-fuentesOrderIntermittencyPressure2013} suggest that the interaction between active and passive stresses becomes more complex near perfect order.

In this work, we further investigate this limit. Using particle-based simulations and coarse-graining the resulting spatially varying stress tensor fields, we show that the picture of an excess stress perpendicular to the growth direction required to stabilize alignment breaks down. Instead, we find that the system can spontaneously organize into states with a long-term average stress anisotropy which is inverted, i.e., showing excess stress in the direction of growth. Starting from asymptotic theories for perfect alignment, we show that the resulting stress tensor is consistent with a decoupling of active and passive stresses, which persists even in the case of open channels, where confinement is not provided by perfectly straight walls but by the fluctuating growing colony itself.

\section{Methods}
\subsection{Agent-Based Modeling}
We use an agent-based model of 
compressible rod-like cells that grow in length and divide.
Each rod is comprised of a rectangular fixed-width body
with half-circle caps at both ends. Over time, the
total cell length increases linearly to $\lmax$, twice its starting
value, and then divides into two daughter cells with 
identical orientation $\varphi$ and randomly drawn growth rate
from a uniform distribution $\gamma \in \left[\frac{3}{4}, \frac{5}{4}\right]$.
Neighbouring rods interact mechanically only and repel one another
according to \emph{Hertzian} repulsion law\cs\cite{hertzUeberBeruhrungFester1882}
\begin{align}
  {\bf F}({\bf d}) =
  \left\{
  \begin{array}{cc}
    \frac{Y}{2}\sqrt{\frac{R}{2}} \left( 2R - |{\bf d}|\right)^{\frac{3}{2}}\hat{\bf d}, & |{\bf d}| \le 2 R \\
    0, & |{\bf d}| > 2 R
  \end{array}
   \right.
   \label{eq:hertz}
\end{align}
as illustrated in Fig.~\ref{pan:hertz}, where ${\bf d}$ is the shortest connection between the two center lines. 
The simulations are set in the over-damped limit, such that the velocity of the $i$th particle ${\bf v}_i=\bmu_i {\bf F}_i$, where ${\bf F}_i$ is the total force on particle $i$ and $\bmu_i$ is an anisotropic mobility tensor.
The model is very similar to other models of dividing rods in the literature\cs\cite{volfson_biomechanical_2008, boyerBucklingInstabilityOrdered2011, orozco-fuentesOrderIntermittencyPressure2013, youConfinementinducedSelforganizationGrowing2021} and builds on physical intuition, hence, we refer the reader to the Supplementary Material for additional details.

\begin{figure}[ht]
    \includegraphics[width=\columnwidth]{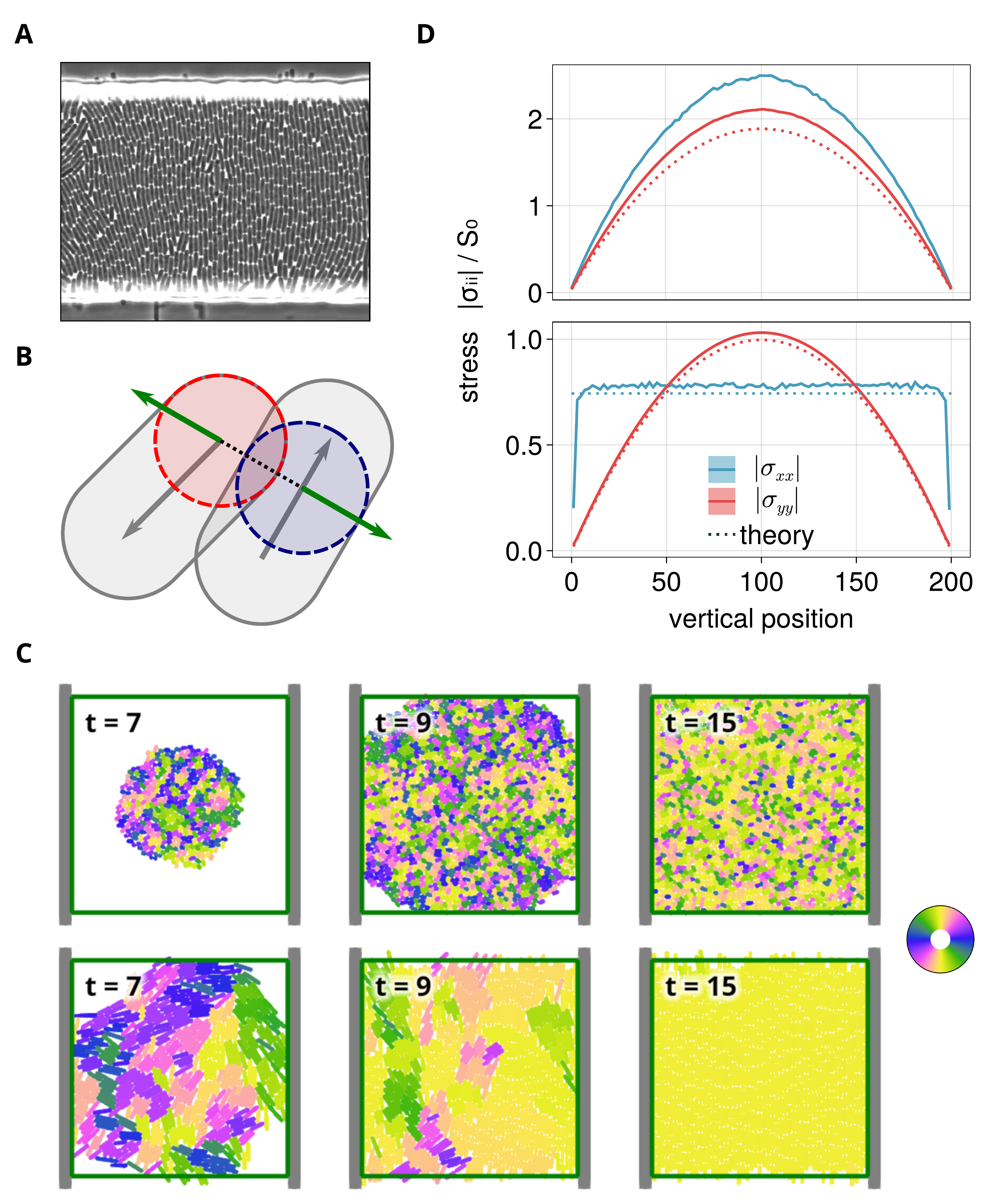}
    \mycaption{\panel{pan:experiment} Experimental observation of growing \textit{E. coli} exhibiting alignment when confined in a quasi-2D microfluidic trap. The openings of the trap are at the top and bottom (bright refracting areas), the confining walls are towards the left and the right (outside the area shown here).
    \panel{pan:hertz}~Illustration of Hertz-based repulsion force law.
   \panel{pan:demo}~System configurations in a $50\times50$ channel after 7, 9, and 15 cell generations for parameter $\lmax=2$ (top row) and $\lmax=6$ (bottom row). 
   Colouring indicates orientation as defined in the colour wheel.
    \panel{pan:demo_profiles} Cuts along the length of a simulation channel with height/length 200 and filled with $\lmax=2$ (top) and $\lmax=6$ (bottom) growing rods. Theoretical predictions (dashed) accompany simulation results.}
    \label{fig:fig1}
\end{figure}

\subsubsection{Confinement}
\label{sec:confinement}
We consider a two-dimensional channel with outlets on the two opposing sides in the $y$ direction and confinement on two sides in the $x$ direction.
Particles are removed once their center point crosses the outlet boundaries.
Confinement in $x$ direction is modeled by walls that exert Hertzian forces as in Eq.~\eqref{eq:hertz} on the cells, where ${\bf d}$ is now the shortest connection from the center line of the cell to the wall.
Two examples of such a channel simulations on a $50\times50$ unit domain are shown in Fig.~\ref{pan:demo}. In our actual investigation, we will use domains of size $200\times200$. All lengths are given in multiples of the rod width, which is kept constant throughout this work and measurements are taken after discarding transient dynamics.

\subsection{Extraction of Continuous Fields}
To study the emergent dynamics at greater length scales, we compute approximate continuum fields of the orientational order parameter and stress tensor on a fine rectangular grid. We write for the former%
\begin{align}
  \Xi(\mv r) = \frac{1}{Z(\mv r)} \sum_{k} \exp(2i\varphi_k) a_{(k,\Delta V)}  \label{eq:complexorder}
\end{align}
where $a_{(k,\Delta V)}$ is the overlapping area of particle $k$ and small sub volume $\Delta V$ around point $\mv r$ and the normalization factor $Z(\mv r) = \sum_k a_{(k,\Delta V)}$ corresponds to the total overlap area of all cells with $\Delta V$.
The average orientation is given by%
\begin{align*}
  \phi(\mv r) = \frac{\text{arg}(\Xi(\mv r))}{2}
\end{align*}%
where $\text{arg}$ refers to the angle in the complex plane and we divide by two to account for the nematic symmetry.
The scalar order parameter within the same region is given by
\begin{align*}
  \xi(\mv r) = |\Xi(\mv r)|
\end{align*}
where $\xi=1$ corresponds to complete alignment and $\xi\rightarrow0$ to no order.
The stress tensor field is computed as\cs\cite{dasLocalStressPressure2019}
\begin{align}
  \bsigma = - \frac{1}{2\Delta V}\sum_k\sum_{l\neq k}\langle\lambda_{kl} \mv F_{kl}\otimes \mv r_{kl}\rangle \label{eq:stresstensor}
\end{align}
where $\mv F_{kl}$ are the cell-cell interaction forces and $0 \leq \lambda_{kl}\leq 1$ is fractional length of the line segment $\mv r_{kl} = \mv r_k - \mv r_l$ that lies within $\Delta V$. Averages across larger areas or along an entire axis can then be computed as required. Throughout this study, absolute stress values are reported in units of $S_0$, for which we arbitrarily choose the central $|\sigma_{yy}|$ for $\lmax=6$ in a domain with our standard height of 200 length units (see Supplementary Material for an explicit definition of $S_0$).

A useful quantity introduced by You et al. is the normalized stress anisotropy\cs\cite{youConfinementinducedSelforganizationGrowing2021}
\begin{align}
    \Delta \Sigma = \frac{|\sigma_{xx}| - |\sigma_{yy}|}{|\sigma_{xx}|+|\sigma_{yy}|}, \label{eq:anisotropy}
\end{align}
which quantifies the excess stress across the channel, i.e., perpendicular to the walls, in the direction of confinement\cs\footnote{The physical meaning is identical to the definition in Ref.~\citenum{youConfinementinducedSelforganizationGrowing2021} despite the difference in notation, which arises from a different orientation of the channel, swapping $x$ and $y$.}.
\section{Results}
\subsection{Spatially Varying Fields}
We set up the numerical experiments as described above by placing four randomly oriented rods in the center of a $200\times200$ unit channel with confining walls at the sides and let the dynamics evolve for many cell division times.
Examples of this process are shown for a smaller domain and division lengths $\lmax=2$ and $\lmax=6$ in Fig.~\ref{pan:demo} with snapshots taken after 7, 9, and 15 generations, respectively.
An important observation is that preferential alignment is always directed towards the channel outlets. However, the final, highly ordered, columnar structure with $\xi\approx 1$ as in the second example is only attained for $\lmax \gtrsim 4$, whereas for $\lmax \lesssim 4$, the steady-state is characterized by imperfect order with $\xi<1$.

For rods  with $\lmax = 2$ (resulting in $\xi\approx 0.5$),  we resolve the stress field of the final steady state along the channel. We find that both $|\sigma_{yy}|$ and $|\sigma_{xx}|$ take on a parabolic profile with $|\sigma_{xx}|$ being consistently larger as shown in Fig.~\ref{pan:demo_profiles} (top). The system is therefore, both locally and globally, characterized by $\Delta \Sigma > 0$, which has previously been observed as typical\cs\cite{youConfinementinducedSelforganizationGrowing2021}, along with the intuitive explanation that stress is more difficult to dissipate in the confined direction and  needed to generate and maintain a preferential alignment towards the channel outlets.

However, cells dividing at a greater maximal length $\lmax=6$ (resulting in $\xi\approx 1$) exhibit a different behaviour as shown in Fig.~\ref{pan:demo_profiles} (bottom): Here, measurements of the stress tensor along a central cut through the channel show that, while the vertical stress $|\sigma_{yy}|$ follows the familiar parabolic profile, the confined horizontal stress locks onto a constant value in large parts of the domain.
Surprisingly, this even leads to an inversion of the stress anisotropy $\Delta\Sigma$ in the colony center, with the horizontal stress $|\sigma_{xx}|$ remaining at lower values than the vertical stress $|\sigma_{yy}|$.

Observing near uniform values of $|\sigma_{xx}|$ while $|\sigma_{yy}|$ changes continuously indicates a decoupling of the stresses made possible by perfect ordering inside the channel.  
We will approach the study of this (de-)coupling from two angles: We begin by deriving limiting theories which reveal the physical origin of $|\sigma_{yy}|$ and $|\sigma_{xx}|$ by predicting them separately. We then continue numerically by modifying the geometry, primarily impacting constraints on the horizontal stress.

\subsection{Column Theory for Active Stress}
A colony of our model rods has a well-defined and spatially homogeneous distribution of cell age $g \in [0,1)$ and growth rate $\gamma \in \left[\frac{3}{4},\frac{5}{4}\right]$ described by
\begin{align}
  p(g,\gamma) = \frac{2\chi}{\gamma}\exp(-\frac{\chi g}{2\gamma})\label{eq:colonydistr}
\end{align}
with a normalization parameter $\chi$ and averages
\begin{align*}
  \langle\gamma\rangle \approx\, 0.985\,\quad\text{and}\quad
  \langle g\rangle \approx\, 0.441.%
\end{align*}
With this distribution at hand, we predict the parabolic stress profile using the emergent and effectively one-dimensional columnar structures.
In this limit, we neglect all transverse (horizontal) dynamics and derive an expression for the $|\sigma_{yy}(y)|$ stress profile within the colony. Also assuming incompressibility, a generic dry continuum $\partial_t{\rho} + \nabla\cdot(v\rho) = \alpha\rho$ for the cell density $\rho$ with an effective growth rate $\alpha$ immediately leads to the condition $\nabla\cdot v=\alpha$ for the steady state, or $\partial_y{v_y} = \alpha$ for our one-dimensional consideration.
The effective rate $\alpha$ at which line density is produced in our model can be calculated from \cref{eq:colonydistr}, such that
\begin{align}
  \pdv{v_y}{y} = \frac{1}{\langle l\rangle} \int  \gamma\,\Delta l\, p(g,\gamma)  \dd{g} \dd{\gamma} \,=\, \frac{\langle\gamma\rangle\Delta l}{\langle l\rangle} \,=\, \frac{\langle\gamma\rangle}{\langle 1+g\rangle}\nonumber
\end{align}
where $\Delta l = \frac{\lmax}{2}$ is the length (including caps) by which a cell grows between divisions and $\langle l\rangle = \Delta l\langle 1+g\rangle$ is the average length occupied by a cell.
By defining $y=0$ as the center of the colony with $v_y(y=0) = 0$, the velocity profile therefore becomes
\begin{align}
  v_y(y) = \frac{\langle\gamma\rangle}{\langle1+g\rangle} y = \mu \pdv{\sigma_{yy}}{y} \label{eq:stressgrad}
\end{align}
where the second equality represents force balance and $\mu$ is an effective mobility that accounts for substrate friction, which, in general, could be a function of density and therefore of $y$. Again using the incompressibility assumption, it is computed as
\begin{align}
  \mu = \frac{\langle l\rangle}{\left\langle\mu_{||}^{-1}\right\rangle} \label{eq:mobility}
\end{align}
where, in the agent-based model, the mobility parallel to the symmetry axis $\mu_{||}$ is a function of the cell aspect ratio $\frac{l}{2R}$ with the cell radius $R$, as defined in the appendix model description, and the average can be computed numerically. With the boundary condition that the stress vanishes at the outlets, $\sigma_{yy}(y_{\max}) = 0$, we integrate \cref{eq:stressgrad} and obtain a parabolic stress profile
\begin{align*}
  \sigma_{yy}(y) =& \frac{\langle\gamma\rangle}{\mu\langle 1+g\rangle}\frac{(y_{\max} - y)^2}{2}
\end{align*}
with the maximum stress at the center.
This result is used in Fig.~\ref{pan:demo_profiles} to predict the stress profile and matches closely for the case of $\lmax=6$ while significantly underestimating the central stress of $\lmax=2$ due to compression effects neglected in the theory. 
A more detailed comparison is shown in Fig.~\ref{pan:central_y_theory}, where the maximal stress $|\sigma_{yy}|$ is computed for a range of division length values $\lmax$.
The theory matches numerical results for long cells, the limit for which the theory was derived, as it yields the near-perfectly ordered quasi-columnar structure. For shorter cells, disorder and compression becomes more relevant, causing the theory to underestimate the measured stresses. We conducted additional simulations with a single one-dimensional colum of cells while freezing the  orientational degree of freedom (grey line in Fig.~\ref{pan:central_y_theory}). The fact that this measurement matches the theory even more closely confirms that a large part of the deviation is caused by transverse compression, which effectively increases the density and thereby also changes the mobility.

It may initially seem surprising that the theory approximately captures $|\sigma_{yy}|$ even at small division lengths $\lmax\lesssim 4$, where the system self-organizes into a weakly ordered state far away from a columnar structure assumed above. However, on a mean-field level, the starting point of our incompressible theory, including $v_x=0$, $\partial_yv_y=\alpha$ and the distribution of cell ages, holds in the disordered system as well. Therefore, the theory can also be viewed as an approximation to the disordered 2D scenario. An additional inaccuracy, besides the violations of the incompressibility assumption explained above, then arises from the anisotropic mobility of the particles, which changes the estimate of $\mu$ in \cref{eq:mobility}.

\subsection{Passive Stress Theory}
To determine whether active and passive stresses indeed decouple in the highly ordered state, we calculate the expected $|\sigma_{xx}|$ arising only from the passive repulsion between neighbouring columns.
This is possible in such a state since the overlap $\Delta<2R$ between neighboring columns can be calculated from the number of columns, the channel width and the width $2R$ of a single cell. Parallel cells at distances $\Delta$ from one another then exert Hertzian repulsion forces of the form
\begin{align}
  f_x = Y\sqrt{\frac{R}{8}} \left(1 - \frac{\Delta}{2R}\right)^{\flatfrac{3}{2}}
\end{align}
on each other.

\begin{figure}
    \includegraphics[width=\columnwidth] {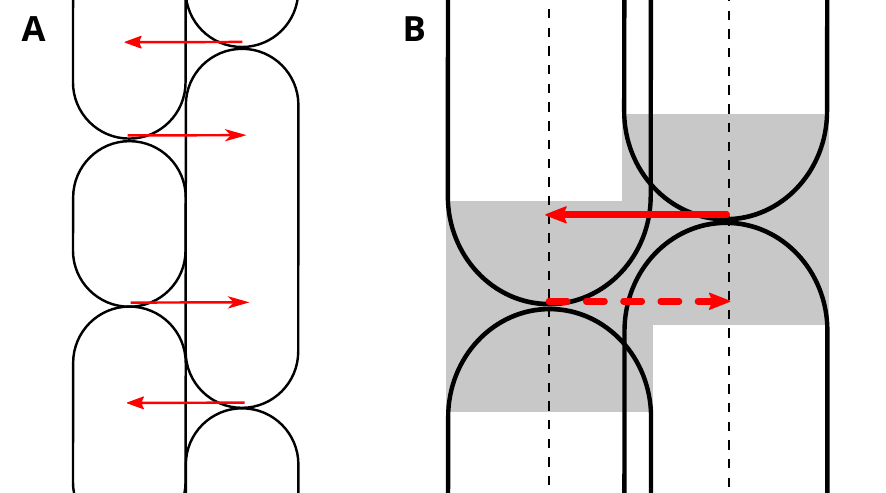}
    \mycaption{\panel{pan:illust} Illustration of counting argument for passive stress theory.
    \panel{pan:detailview} Detailed view of missing interaction.}
    \label{fig:graphicalapproach}
\end{figure}

We then consider the line density of these force vectors along the vertical axis:
Each cell pair has at most one interaction, but depending on the cell length, which ranges between $\lmax/2$ and $\lmax$, and configuration, some cells may interact with up to three cells on each side, as illustrated in \cref{fig:graphicalapproach}.

Considering two adjacent columns, we can count the number of unique touching cell pairs. Every new cell in the left column has one interaction partner on the right and likewise every new
cell on the right interacts with the one to its left as is indicated by red arrows
in Fig.~\ref{pan:illust}. 
This scheme of counting captures all interactions and we write
\begin{align*}
  \sigma_{xx, passive}(\Delta) =  \frac{2}{\langle l\rangle}Y\sqrt{\frac{R}{8}}\left(1 - \frac{\Delta}{2R}\right)^{\flatfrac{3}{2}} \left(1-\frac{1}{2}\frac{4R}{\langle l\rangle}\right)
\end{align*}
since on average there is a new cell every $\langle l\rangle$ in each column. The rightmost term represents a correction factor:
As illustrated in Fig.~\ref{pan:detailview}, whenever the cell caps are too close to those in the other column, one of the two potential interactions is lost. 
Assuming there are no long-range correlations between the neighbouring columns, two cell cap regions are expected to overlap with probability $p = 4R\langle l\rangle^{-1}$.

A first comparison with numerical estimates is shown in Fig.~\ref{pan:demo_profiles}~(bottom) where a dashed line indicates the expected configurational stress $|\sigma_{xx}|$ for 205 columns in the 200 unit wide channel using $\lmax=6$.
The measured value itself is matched well by our theory, again with a slight underestimation due to neglected compression along the columns, which increases the line density of cells and therefore of the interactions.

\subsection{Statistically negative stress anisotropy}
An important input for the prediction is the exact number of columns in the configuration, which is not predetermined by the parameters of the system. Instead, it is an an emergent result of the self-organization process that leads to perfect alignment and, in principle, can also fluctuate over long time scales\cs\cite{orozco-fuentesOrderIntermittencyPressure2013}.
Therefore, the question arises whether the inverted anisotropy $|\sigma_{xx}| < |\sigma_{yy}|$ in Fig.~\ref{pan:demo_profiles}~(bottom) is typical, or, more generally, what the distribution of the number of columns is that the system naturally attains.

To evaluate this in detail, we generated channel simulations with $\lmax=6$ for slightly varying domain width. 
The result is shown in Fig.~\ref{pan:marginal_central_stress}.
The allowed discrete configurational stress values for a fixed number of columns change continuously with increasing domain width (dashed lines). For each value of the domain width, 20 independent initial conditions lead to a spectrum of observed of column numbers, whose stress values show excellent agreement with the corresponding discrete levels. While a few simulation runs generated a state with $|\sigma_{xx}| > |\sigma_{yy}|$, the vast majority exhibit $\Delta\Sigma < 0$. This is further emphasized by the marginal distributions shown in Fig.~\ref{pan:marginal_central_stress}.
\begin{figure}[!h]
    \includegraphics[width=\columnwidth]{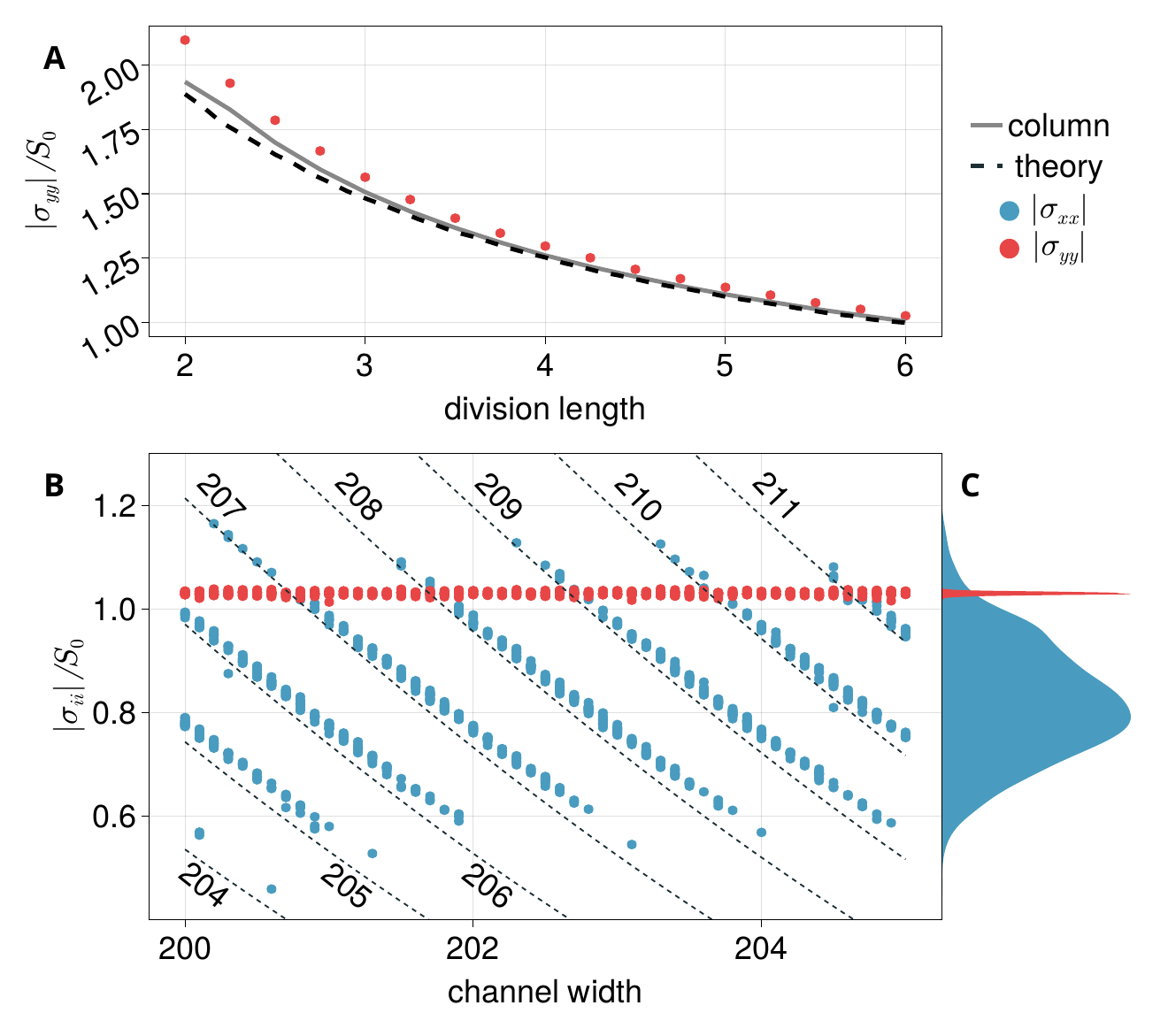}
    \mycaption{
    Evaluation of theoretical predictions. \panel{pan:central_y_theory} Central y-stress in channels (blue) with theory (dashed) and a simplified 1D column simulation (grey).
    \panel{pan:central_stat_stress} Stationary central stress field components for cells with $\lmax=6$. $|\sigma_{xx}|$ and $|\sigma_{yy}|$ are measured in a channel-wide 10 unit tall slice (as illustrated in \supprefc) and averaged over five generations. Predictions for various column numbers are shown as dashed lines.
    \panel{pan:marginal_central_stress} Marginal distributions of points in Figure~\ref{pan:central_stat_stress}.
    }
    \label{fig:fig_widthscan}
\end{figure}

\subsection{Open Domains}

\begin{figure}
    \includegraphics[width=\columnwidth]{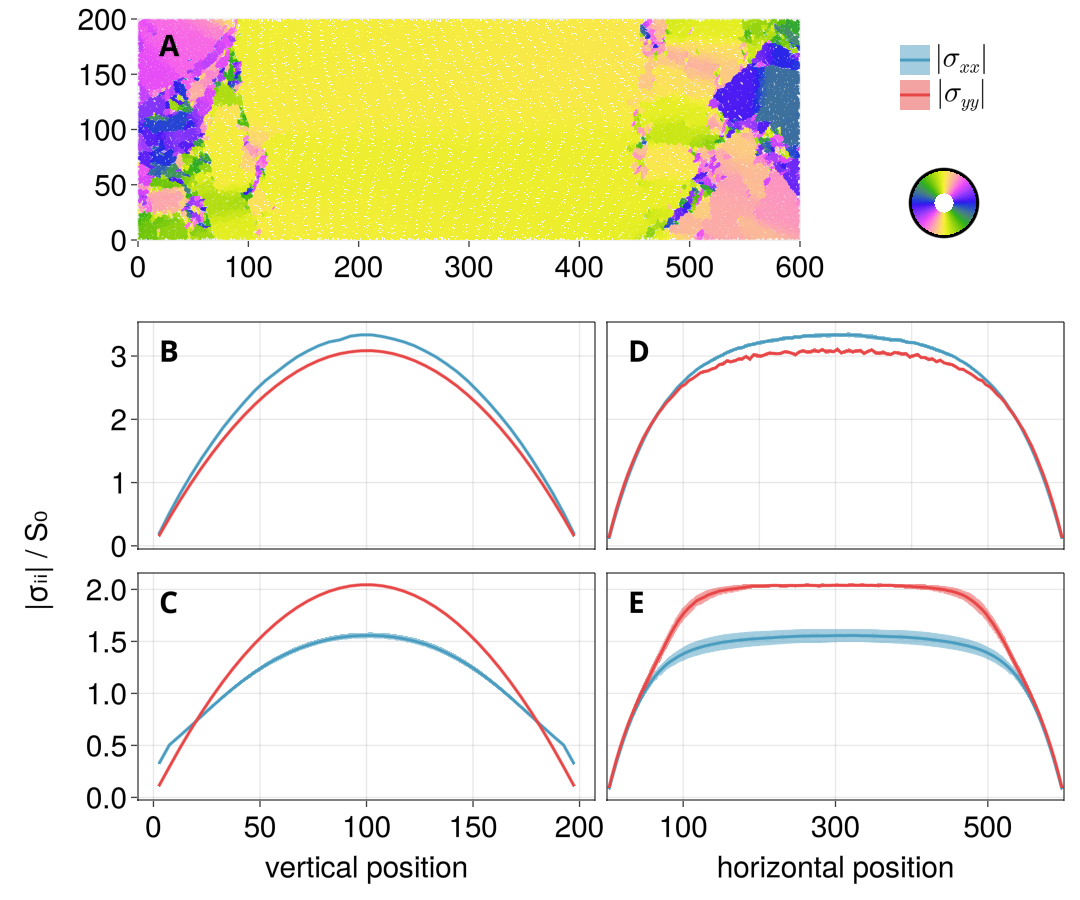}
    \mycaption{
    Open rectangular domains. 
    \panel{pan:open_snap} A snapshot of a $600\times 200$ unit open domain filled with $\lmax=6$ growing rods.
    Colour indicates rod orientation as illustrated in the colour wheel. 
    \panels{pan:profb}{pan:profc,pan:profd}{pan:profe} Central vertical and horizontal cuts through the domain measuring the local stress tensor coefficients for $\lmax=3$ (top) and longer $\lmax=6$ (bottom) cells.
    Approximations were computed as illustrated in \supprefc using 10 initial conditions for $\lmax=3$ and 210 initial conditions for $\lmax=6$ to better capture increased fluctuations.
    }
    \label{fig:fig3}
\end{figure}

The channel geometry used so far is in many ways special: Not only does it impose a preferential direction for growth (or, in fact, allows for only one possible flow direction), it also has perfectly straight walls which ensure that the same amount of space is available to each column as cells are pushed along the channel. Furthermore, the walls do not introduce any perturbations. As we have seen, the passive horizontal stress is purely determined by these boundary conditions and is therefore also constant along the channel. Given that the growth-induced active stress nevertheless drives the self-organization process that selects a certain number of columns, it may seem as if the special properties of the channel confinement alone are responsible for the decoupling of the two stresses and other unusual features of this state, such as the negative stress anisotropy $\Delta\Sigma<0$.

To find out whether this is the case and better understand the origin of emergent effects, we carry out simulations in rectangular domains that have outlets on all four sides.
Open domains provide a weaker constraint on the dynamics as they permit, \emph{a priori}, flow in all directions.
Still, for \emph{non-square} domains, we find preferential alignment between the long opposing sides,
as shown in Fig.~\ref{pan:open_snap}. 
From the initial unordered phase, the dynamics produce a highly ordered central region supported by disordered sides that generate the required compression.
This is a non-trivial result in itself and highlights the system's natural tendency to generate order.
However, due to the complex dynamics in the side regions, the ordered phase may spontaneously become unstable leading to a macroscopic buckling event before the ordered state is eventually reestablished. An example of this is shown in \supprefd.

Measurements of the stress tensor fields are shown in Fig.~\ref{pan:profb}-\ref{pan:profe}. 
Analogous to Fig.~\ref{pan:demo_profiles}, these plots display stress profiles as measured along a vertical cut through the domain center (Fig.~\ref{pan:profb} and \ref{pan:profc}) and, in addition, the same along the horizontal direction (Fig.~\ref{pan:profd} and \ref{pan:profe}).
For short cells, the behaviour along the vertical shown in Fig.~\ref{pan:profb} is the same as in channels and along the horizontal axis (Fig.~\ref{pan:profd}) we find a profile that flattens towards the center. 
In the case of longer cells, as displayed in Fig.~\ref{pan:profc} and \ref{pan:profe}, the behaviour deviates from the channel simulations.
Most importantly, the open domains do not allow for spatially homogeneous passive horizontal stress.
Instead, horizontal compression is actively generated near the domain sides which can be seen from the steep increase in stress along the horizontal cut in Fig.~\ref{pan:profe}.
Still, the dynamics in the center are quite similar to the channel, as no horizontal motion is apparent and no active stress is generated horizontally due to vertical nematic ordering (compare snapshot in Fig.~\ref{pan:open_snap} and the flat part of the horizontal profile Fig.~\ref{pan:profe}).
Therefore, the horizontal stress in the central region can again be expected to be mostly of passive origin, due to the compression of vertical cell columns. However, the fact that the horizontal stress varies vertically in Fig.~\ref{pan:profc} (in contrast to Fig.~\ref{pan:demo_profiles} bottom) means that the space that columns occupy is not constant as in the channel, but increases from the middle towards top and bottom.
Therefore, the open geometry successfully removed this artificial constraint imposed by the channel walls and, at the same time, introduced perturbations from the disordered sides, lending additional weight to the observation, that, nevertheless, clearly a negative anisotropy $|\sigma_{xx}| < |\sigma_{yy}|$ is observed on average.

\subsection{Length dependency}
\begin{figure}[!h]
    \includegraphics[width=\columnwidth]{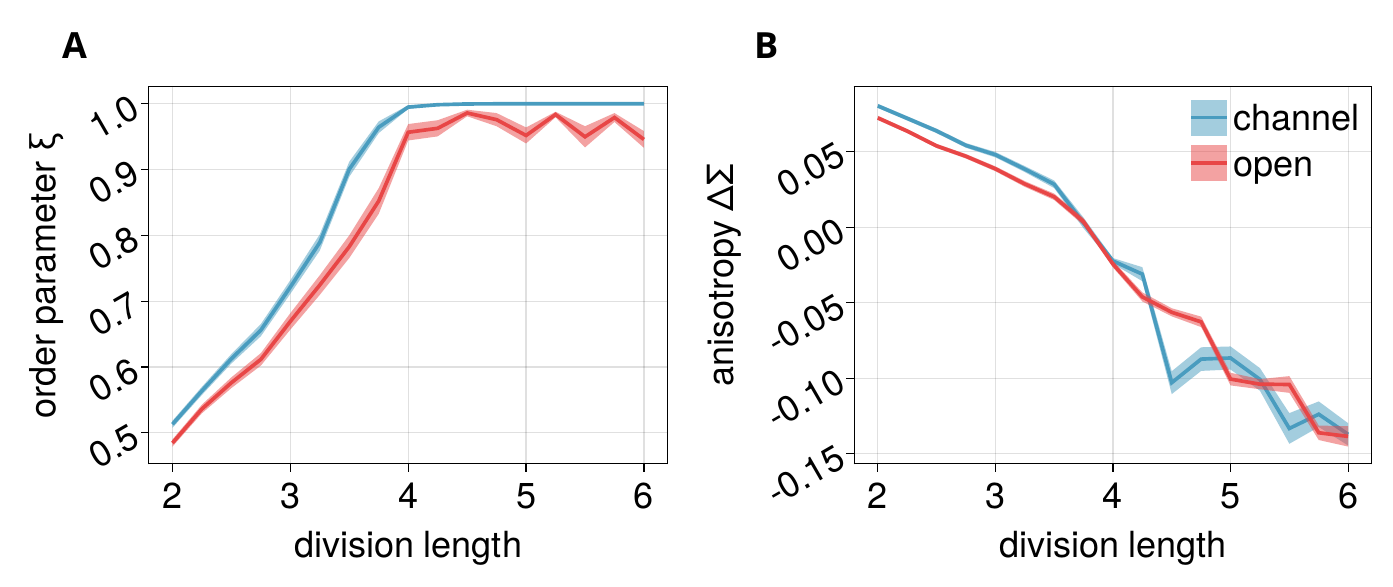}
    \mycaption{Direct comparisons of channels and open domains under varying cell division length $\lmax$. 
    \panel{pan:lengthscan_order} Order parameter in a central $100\times 20$ region as illustrated in \supprefc averaged over time and initial conditions.
    \panel{pan:lengthscan_anisotropy} The similarly averaged central anisotropy for channels and open domains.}
    \label{fig:lengthscan}
\end{figure}

So far, we have only considered exemplary cell division lengths $\lmax$ to understand the qualitatively different dynamics of short and long cells. It is therefore natural to ask how the results compare quantitatively across the entire $\lmax$ range and between channels and open domains.
A finely sampled length scan is summarized in \cref{fig:lengthscan}, displaying the average order parameter $\xi$ (Fig.~\ref{pan:lengthscan_order}) and stress anisotropy $\Delta\Sigma$ (Fig.~\ref{pan:lengthscan_anisotropy}) for both geometries.
These values are computed in a central  $100\times 20$ box around the center of the domains.
Averaging is done over 150 instantaneous measurements sampled from 15 generations of time evolution and additionally over 10 independent initial conditions.

While always being slightly more disordered, the dynamics in the open domain centre closely resembles that of the channels across the full parameter range.
In both cases, the order approaches unity for division lengths $\lmax \gtrsim 4$ and this is accompanied by a transition to negative anisotropy $\Delta\Sigma$.
Interestingly, even though the order is slightly reduced in the open domains for all $\lmax$, they exhibit a cleaner relation of $\lmax$ and $\Delta\Sigma$ as shown in Fig.~\ref{pan:lengthscan_anisotropy}. Both effects mostly likely have the same explanation, namely, that the lack of hard boundaries in combination with the larger open domain reduces finite-size effects in the column number.

\section{Conclusion}
We employed agent-based simulations to study the growth of bacterial colonies in confinement.
This topic has seen a range of publications in recent years, but a careful analysis of the interaction forces and the corresponding spatially resolved stress fields has revealed previously unreported phenomena.

To understand the dynamics from a theoretical perspective, we focused on the limiting case of perfect ordering.
In this limit, it was possible to predict the (active) vertical stress field.
It gives reasonable predictions even for short cells that do not generate the column structure used in the derivation.

The column structure also allowed for an estimation of the
configurational horizontal stress. Comparison with simulation data confirmed 
that, in the column structure, the horizontal stress is indeed largely
passive with the theoretical prediction matching measurements up to a few percent.

Many of the qualitative features of these states were also found in simulations on open rectangular domains, which revealed that, even without hard and flat walls, near-perfect order and strong negative stress anisotropy may emerge.
This is an important result, as it indicates that neither strong ordering nor the negative stress anisotropy were an artifact of the particular confining channel, but rather a natural emergent property of the dynamics. 
It provides a compelling argument that the ordered state can indeed be stable in the presence of a (limited) stress anisotropy inversion.
Away from perfect order, there is a tight coupling of the stress field components and order, as any imbalance in the interaction forces causes microscopic reorientation away from the direction of strongest compression.
This feeds back to the stresses, both through relaxation of the existing stress, as well as by redirecting the active stress of growth.
However, as the order parameter approaches unity, compression along the principal axes of the orientation field ceases to generate torques on the rods involved and the components of the stress tensor decouple.

An interesting connection of our work to that of You et al.\cs\cite{you_geometry_2018} is that, in our open domains, we observe the characteristic formation and breakup of microdomains. The rectangular geometry formed by outlets is sufficient to drive the described ordering both instantaneously and in the long-time average. However, at intermediate timescales, stochastic buckling events are observed that break up even the macroscopically ordered domain in the middle region into smaller fragments. 
The frequency of these events depends strongly on the relative scales of the total domain size versus typical microdomains which themselves depend on the division length $\lmax$.
This effect required the use of many independent initial conditions to produce Fig.~\ref{pan:profe} without significant fluctuations in the $|\sigma_{xx}|$ profile.
The statistics of these events in open domains and the connection to the smectic properties of the column structure seem to warrant more detailed future investigations (as done for the channel geometry\cs\cite{orozco-fuentesOrderIntermittencyPressure2013}).

Dell'Arciprete et al.\cs\cite{dellarcipreteGrowingBacterialColony2018} explain the underlying alignment mechanism of the emergent dynamics using torques experienced by rods in shear flow, similar to recent observations in other geometries\cs\cite{basaranOrientationalOrderInward2022}. 
You et al.\cs\cite{youConfinementinducedSelforganizationGrowing2021} on the other hand attribute this to an anisotropy in the stress tensor.
One could, of course, argue that these two are not very different at all, given that the stress field creates flows in the first place.
However, our work adds to this discussion. We, unexpectedly, found an inversion of stress anisotropy in the strongly ordered limit, even in open domains.
This is not consistent with a purely stress based coupling. 
On the other hand, the observed stochastic collapsing of the ordered structures in the open domains can not be explained using flow based arguments only, as it lacks a destabilizing mechanism for this case.
A theory combining both effects may be a promising avenue for capturing all observed phenomena.

In this work, we presented a detailed analysis of a computational model of growing rod-shaped particles and discovered states in which activity-induced stresses, passive nematic properties, volume exclusion and confinement can interact in novel ways and yield strongly anisotropic systems.
Despite these emergent intricacies, our model merely captures one specific physical aspect of reality when it comes to the growth of bacteria in colonies or biofilms.
It would be interesting to elucidate what role the effects discovered here can play among the complex processes that characterize natural systems. This includes nutrient distribution\cs\cite{bittihn_genetically_2020,Wang2017,martinezcalvoRoughening2022}, which leads to gradients of activity and thereby represents another opportunity for the passive properties of the system to become important, but also extends to structure formation in three dimensions\cs\cite{Wang2017,Hartmann3DBiofilm2019,martinezcalvoRoughening2022}, interactions with other types of activity such as motility and chemical signal production\cs\cite{mooreottBiophysicalThreshold2022} and different kinds of confinement\cs\cite{fortuneElasticConfinement2022}.
In addition, it may prove rewarding to look for even more general physical principles by exploring the similarities the system exhibits to dense systems of active polymers and filaments\cs\cite{winklerActiveFilaments2020,abbaspourActiveFilaments2022} due to its columnar structure.

\section*{Conflicts of interest}
There are no conflicts to declare.

\section*{Acknowledgements}
We thank Benoît Mahault and Yoav G. Pollack for stimulating discussions as well as Jeff Hasty at the University of California San Diego for the permission to use the experimental picture in Fig.~\ref{pan:experiment}.

\end{document}